\documentclass[journal]{IEEEtran}
\ifCLASSINFOpdf
   \usepackage[pdftex]{graphicx}
  \graphicspath{{./}}
   \DeclareGraphicsExtensions{.pdf,.jpeg,.png}
\else
  \graphicspath{{./}}
   \DeclareGraphicsExtensions{.eps}
\fi
\usepackage{amsmath}
\usepackage{color}
\usepackage{caption}
\usepackage{subcaption}
\usepackage{wrapfig}
\usepackage{float}
\restylefloat{figure}
\usepackage{epstopdf}

\begin{document}
%
\title{Non-parametric Impedance based Stability and Controller Bandwidth Extraction from Impedance Measurements of HVDC-connected Wind Farms}

%
%

\author{Mohammad~Amin,~\IEEEmembership{Member,~IEEE,}
        and~Marta~Molinas,~\IEEEmembership{Member,~IEEE}
\thanks{The authors are with the Department of Engineering Cybernetics, Norwegian University of Science and Technology, Trondheim-7491, Norway. (e-mail:mohammad.amin@ntnu.no; marta.molinas@ntnu.no). }}

\maketitle

\begin{abstract}
 
Impedance measurements have been widely used with the Nyquist plot to estimate the stability of interconnected power systems. Being a black-box method for equivalent and aggregated impedance estimation, its use for the identification of sub-components bandwidth is not a straightforward task.  This paper proposes a simple method that will enable to identify the specific part of the equivalent impedance (e.g. controller's bandwidth) that has major impact on the stability of the system.   
For doing that, the paper analyses the stability of an interconnected system of wind farms and high voltage dc (HVDC) transmission system. The impedance frequency responses of the wind farms and HVDC system from the ac collection point are measured and it is shown by the method proposed in this paper, which controller has major impact in the observed oscillation. A mitigation technique is proposed based on re-tuning of the critical controller bandwidth of the interconnected converters. The method suggested can reveal the internal controllers' dynamics of the wind farm from the measured impedance combined with an analytical expression of the impedance and a transfer function identity when no information about the controllers is provided by the vendors due to confidentiality and industry secrecy.

\end{abstract}


%
\IEEEpeerreviewmaketitle

\section{Introduction}

 The impedance-based analysis, first presented in \cite{RDMiddlebrook1976}, has proven to be a useful tool to assess the stability of  interconnected system of wind farms and high voltage dc (HVDC) transmission system  \cite{HanchaoLiu2014IEEEJESTPE}-\cite{MAminJESTPE2016}. The impedance measurement is the basis in the impedance-based approach for the Nyquist criterion to estimate the  stability \cite{JianSun2011}. In order to apply this approach, deriving the analytical impedance model of inverters is the prerequisite and the correctness of the analytical impedance model is verified by numerical simulations or experiments \cite{CespedesM2011IEEETransPE}-\cite{BoWen2016IEEETransPE2}. The analytical impedance model is a continuous transfer function and the Nyquist criterion is checked on the transfer function of the source-load impedance ratio. To derive the analytical model of the impedances, a detailed modeling of the power electronics converters is required.
However, in the case of wind energy conversion system (WECS), the detailed modeling of the wind turbine generators (WTGs) is generally not be available due to industry secrecy and confidentiality. The WECS is then assumed as a 'black/grey- box' system since no information about the internal control dynamics is available from the vendors. 
Due to this lack of availability, an analytical impedance model for the WECS can not be obtained with good accuracy.
One can argue that it would be enough to measure the impedance of the inverter and then apply the Generalized Nyquist criterion (GNC) on the measured source-load impedance and this can be true for a grid-tied inverter \cite{CespedesM2014IEEETransPE}, however in the case of offshore wind farms application, the system has multiple wind power inverters which are powered by the HVDC system rather than a strong ac grid and the aggregated impedance frequency responses of the wind farms can only be obtained from measurement if the interconnected system of wind farms and HVDC operates stably. In real world application, there is no guarantee that the system will operate stably unless adequate stability measures have been taken before installation. 
Therefore, to ensure stable operation, it is necessary to assess the stability of the interconnected system analytically before connecting to the ac grid \cite{GOKalcon2012IEEETransPS}-\cite{FanLingling2012}. In order to evaluate the stability analytically before connecting to the main ac grid, it is necessary to have a continuous transfer function for the analytical formulation of the impedance model from the black-box approach of the WECS. Moreover, it is necessary to extract the controller dynamics of the inverters to identify participation contribution in the observed oscillation and avoid the interaction phenomena between the controllers of the HVDC rectifier and wind power inverter \cite{HanchaoLiu2014IEEEJESTPE}, \cite{MAminJESTPE2016}.

To the authors knowledge, extraction of controllers' dynamics of an inverter from impedance measurements has not been reported, when information on the controller parameters is not available.
This paper proposes a method for finding the internal controllers' dynamics of the wind power inverter from impedance measurements when the WECS is assumed to be a black box.
The main contribution of the paper to the state-of-the-art is in the following:
\begin{itemize}
    \item The paper presents a novel technique to extract the internal control dynamics of the WECS inverter when the WECS system is assumed to be a 'black box' using system identification techniques.
    \item The paper discusses the role of the ratio between the bandwidths of the phase-lock-loop (PLL) and the HVDC rectifier control-loop, as having an essential role in the root cause of instability and in being a strong factor to be taken into account in the shaping of the impedances to guarantee stability. 
\end{itemize}
 These two aspects of the contribution are discussed in more detail in the following.

\begin{figure*}[t!]
    \centering
    \includegraphics[width=.90\textwidth]{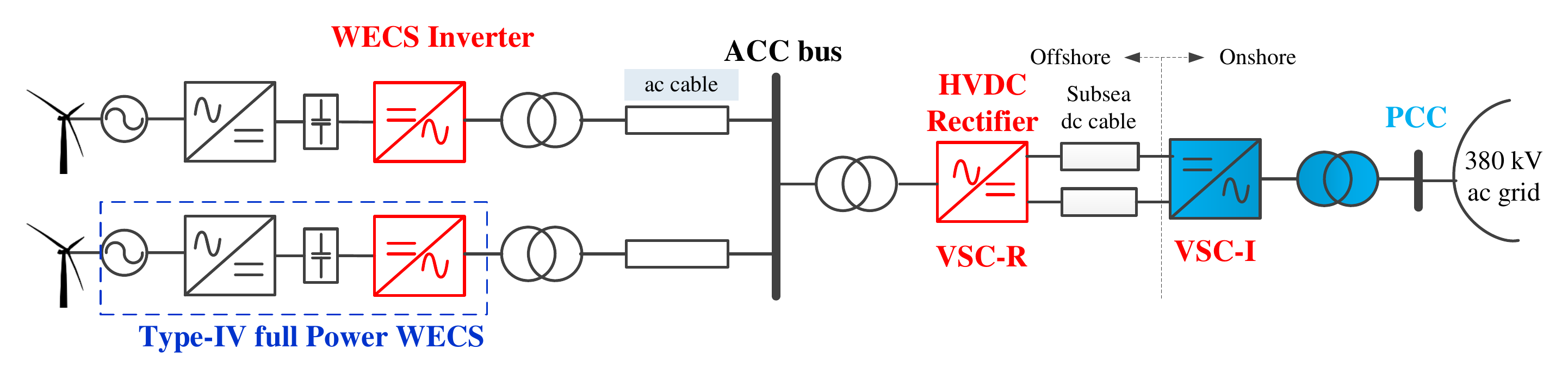}
\caption{Interconnected system of wind farms and VSC-based HVDC system.}
\label{fig:System}
\end{figure*}
\begin{figure}[!t]
\centering
\includegraphics[width=.50\textwidth]{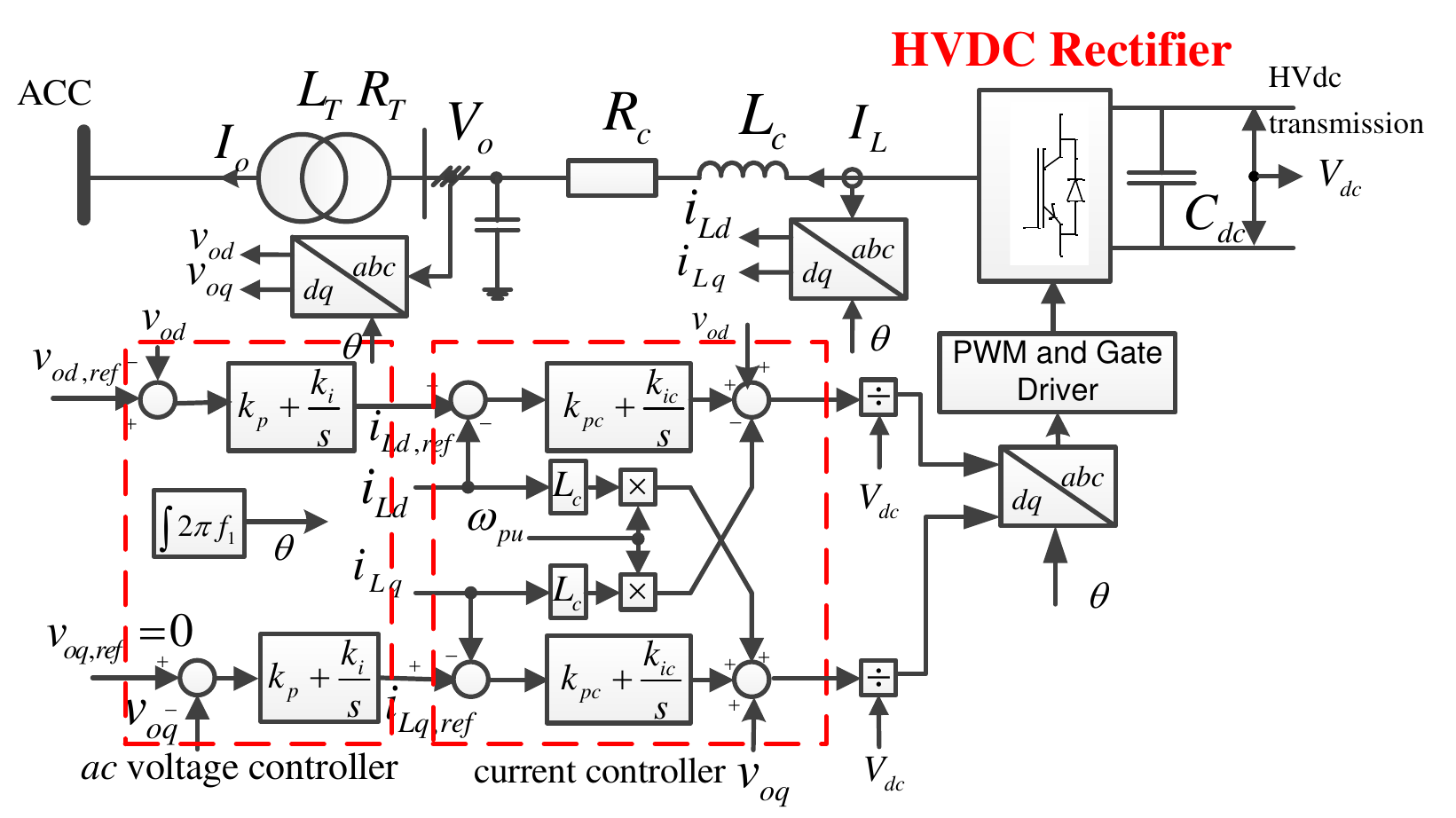}
\caption{Control structure of HVDC rectifier, VSC-R.}
\label{fig:Control_of_VSC_R}
\end{figure}

Since the transfer function of the WECS impedance can not be derived analytically due to lack of detailed modeling information, this paper proposes a method to obtain the aggregated-impedance transfer function from the measured impedance data of a single WTG by using system identification \cite{BWahlbergIEEETransAC1994} and then the impedance-based stability criterion is applied to assess the stability on the aggregated impedance ratio of the interconnected system.
The frequency domain impedance characteristics both for the wind farms and the HVDC rectifier from the offshore ac collection (ACC) point are presented to identify the potential resonance. The Nyquist plots of the impedance ratio of the HVDC converter to the wind farm indicate the potential resonances at low frequency. An unstable case is shown which is originated by the controller interaction even when the control tuning satisfies the standard control tuning independently designed for the converter. 
The method presented in this paper enables to identify which parts from the controllers are participating in the observed oscillatory behaviour and interaction based on the impedance-based method. Moreover, the role of the ratio between the bandwidths of the phase-lock-loop (PLL) and the HVDC rectifier control-loop is highlighted, as having a relevant impact in the stability and in being a strong factor to be taken into account in the shaping of the impedances to guarantee the stability of the system. It has been observed that the source HVDC converter ac voltage control-loop must be ten times higher than the wind power inverter's PLL control loop bandwidth to avoid the instability caused by the controller interaction. Since the wind turbine is assumed to be a black-box and no information on the PLL bandwidth is available, the impedance frequency data from measurement is used to reveal the PLL control-loop bandwidth.

The rest of the paper is organized in the following. Section II discusses the interconnected system configuration and the simulation results. Section III shows the stability analysis of the interconnected system based on the impedance-based method. Section IV presents the method to extract the internal control dynamics of the inverter from the measurement data. In addition, this section presents the interaction analysis between the interconnected converters controllers. Finally the study concludes in Section V.

 

\section{Interconnected System Configuration, Control and Simulation Result}

\subsection{HVDC System Configuration and Control}
The interconnected system under study is depicted in Fig. \ref{fig:System}. The system has two parts. The right part from the ACC bus is the HVDC system and the left part is the wind farms. The HVDC system consists of converter transformers, offshore HVDC rectifier (VSC-R), subsea dc cable, and grid-side onshore HVDC inverter (VSC-I). The VSC-HVDC system has a capacity of 500 MVA equivalent. The VSC-R is connected to the offshore ACC bus through a transformer with same rating as the converter. The VSC-I is connected to the onshore ac grid of 380 kV through a 220/380-kV, 50-Hz, 500-MVA transformer. The HVDC-link dc voltage is 360 kV and the length of the dc line is 100 km.

The VSC-R behaves as a voltage source to the ac terminal and regulates the ac voltage and frequency, while the VSC-I regulates the HVDC-link dc voltage and reactive power. In this work, the focus is to study the interaction between the WECS inverter and the VSC-R, therefore  the detail of modeling and control will be presented only for the VSC-R assuming that the \emph{dc} voltage-controlled converter VSC-I is providing constant dc voltage input to the VSC-R.
The modeling, analysis and the control of the system will be presented in a synchronous reference frame (SRF). The transformation of the three phase quantity from stationary reference frame to SRF is based on the amplitude-invariant Park transformation, with the d-axis aligned with the phase-A voltage vector and q-axis leading the d-axis by 90$^0$.
The electrical circuit and control structure of VSC-R is shown in Fig. \ref{fig:Control_of_VSC_R}. It has inner-loop current control and outer-loop ac voltage control. The converter is providing 50 Hz frequency to the offshore ACC bus.

\begin{figure}[!t]
\centering
\includegraphics[width=.50\textwidth]{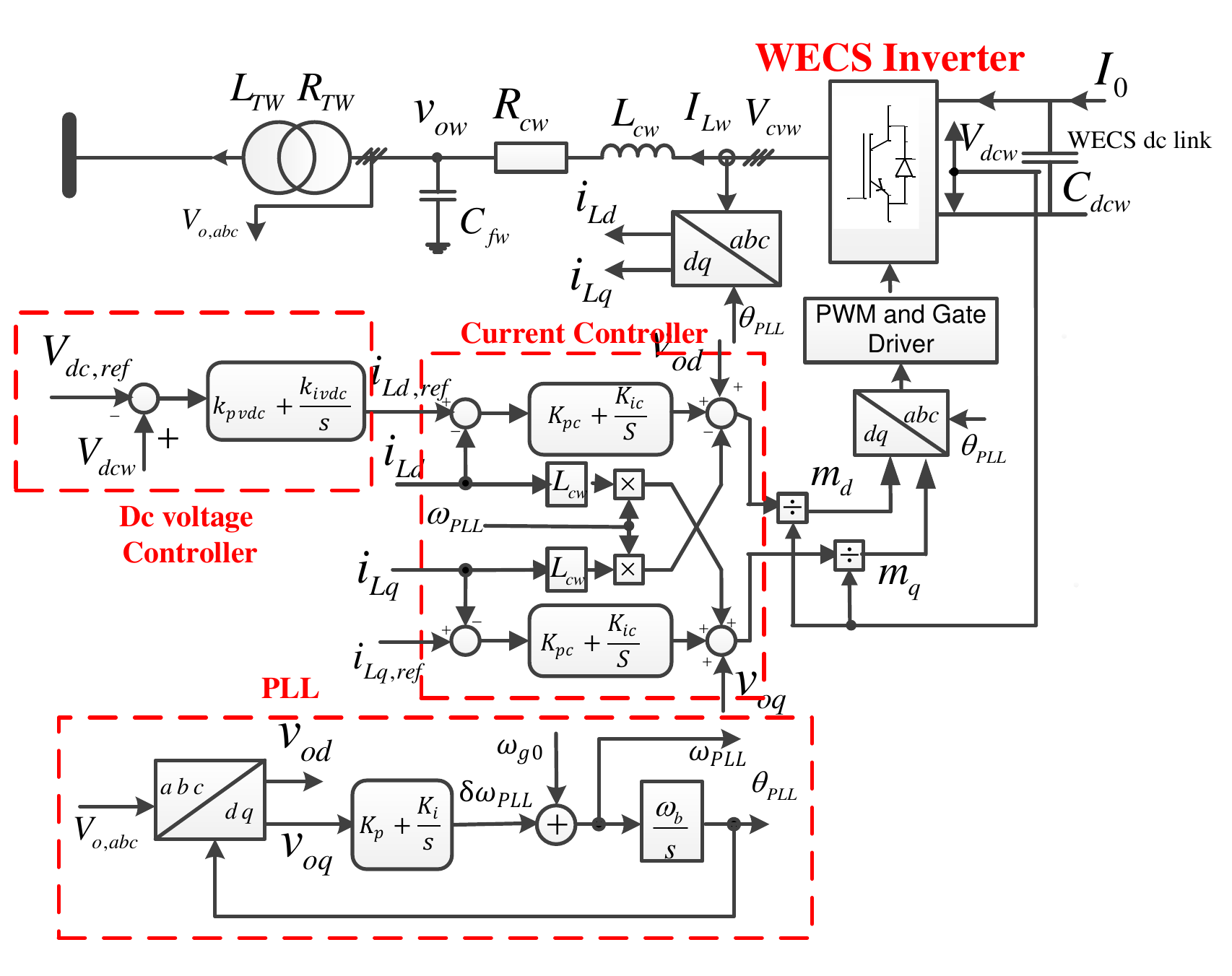}
\caption{Control structure of WECS inverter, WECS-I.}
\label{fig:Control_of_WECS_I}
\end{figure}

\begin{figure}[!t]
\centering
\includegraphics[width=.50\textwidth]{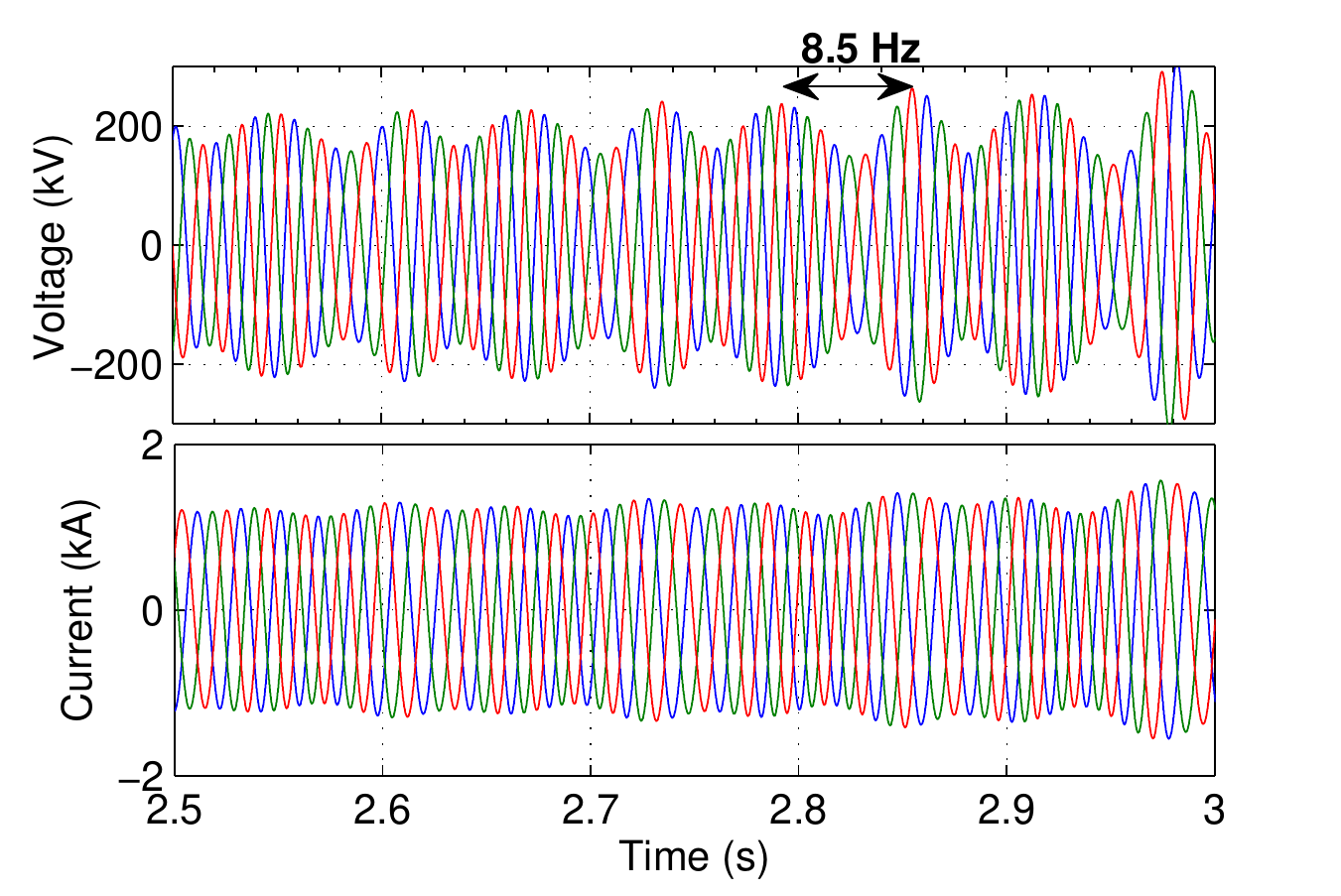}
\caption{Unstable: Three phase voltages and currents at ACC bus.}
\label{fig:VI_ACC_Unstable}
\end{figure}

\subsection{Wind Farms Configuration and Control}
The wind farms shown in Fig. \ref{fig:System} are  connected to the ACC bus through a transformer and undersea cable. The WTGs are assumed to be a type-IV back-to-back WECS. Each wind farm is assumed to have 50 turbines with 3 MW rating each. To simplify the system model, 2x50 turbines are lumped into one unit of WTG with same generation capacity as the wind farms. The generator side VSC (WECS rectifier) regulates the power of the wind turbine based on the predicted wind speed and maximum power point tracking. 
The ac grid side VSC (WECS-I) regulates the dc link voltage of the WECS and reactive power. 
The control structure of the WECS-I is shown in Fig. \ref{fig:Control_of_WECS_I} which has an inner-loop current control in SRF. The d-axis current reference is obtained from the outer-loop dc voltage controller and the q-axis current reference is set according to the reactive power demand. 
A PLL as shown in Fig. \ref{fig:Control_of_WECS_I} is used to track the offshore grid frequency \cite{VKaura1997} and synchronize with the ac collection bus voltage.

\subsection{Simulation Results}
The interconnected system of the wind farm and HVDC transmission system has been implemented in MATLAB/Simulink association with SimPower System Blockset. The electrical circuit parameters of the HVDC system are given in Table \ref{tab:VSC_R}. The current-controller of the VSC-R is tuned at $H_{i,VSC-R}(s)=0.6366+14.25/s$ in pu with 90 degrees phase margin at 400 Hz crossover frequency. The  ac voltage control-loop is tuned at $H_{vac}(s)=0.09+40/s$ with 40 degree phase margin at 80 Hz crossover frequency. The switching frequency of HVDC VSC is 2 kHz. The ac voltage control-loop bandwidth is around 5 times less than the inner-loop current controller and that satisfies the standard bandwidth ratio \cite{LHarneforsIEEETIE2007}. Therefore, the HVDC system is expected to operate stably. 

The electrical circuit parameters of WECS are given in Table \ref{tab:WECS-I}.
The WECS is a 'black box' and no information about the internal control parameter is known. It is assumed that the control-loops of the WECS have been tuned with sufficient phase margin to ensure stable operation. 

\begin{table}[t!]
\caption{The VSC-HVDC system parameters}
\renewcommand{\arraystretch}{1.2}
\label{tab:VSC_R}
\noindent
\centering
    \begin{minipage}{\linewidth} 
    \renewcommand\footnoterule{\vspace*{-5pt}} 
    \begin{center}
   \begin{tabular}{  l  l  l   l }
     \hline
\hline
    Parameter & Value & Parameter & Value \\ \hline
    Rated Power, S$_{b}$ & 500 MVA &  L$_c$ & 0.08 pu\\ \hline
   Rated ac voltage & 220 kV &  R$_c$ & 0.00285 pu \\ \hline  
Trans. inductance & 0.1 pu &  C$_{f}$  & 0.074 pu \\ \hline
Trans. resistance & 0.01 pu & V$_{dc}$ & 360 kV\\ \hline \hline
    \end{tabular}
        \end{center}
    \end{minipage}
\end{table}

\begin{table}[t!]
\caption{Parameter of ACC side WECS VSC}
\renewcommand{\arraystretch}{1.2}
\label{tab:WECS-I}
\noindent
\centering
    \begin{minipage}{\linewidth} 
    \renewcommand\footnoterule{\vspace*{-5pt}} 
    \begin{center}
\begin{tabular}{  l  l  l   l }
    \hline
\hline
    Parameter & Value & Parameter & Value \\ \hline
    Rated Power, S$_{b}$ & 150 MW &  $L_wf$ & 0.12 pu \\ \hline
   Rated ac voltage & 575 V &  $R_wf$ & 0.00285 pu \\ \hline  
Rated dc voltage & 1100 V &  $C_{wf}$  & 0.074 pu \\ \hline
Trans. inductance & 0.04 pu & f & 50 Hz\\ \hline
Trans. resistance & 0.005 pu & $C_{dc}$ & 4 pu\\ \hline
    \end{tabular}
        \end{center}
    \end{minipage}
\end{table}

A time domain simulation has been carried out and the resulting time domain responses are shown in Fig. \ref{fig:VI_ACC_Unstable}. The system is unstable in the time domain simulation even all the parameters and controller satisfies the standard modeling and tuning. As can be seen in Fig. \ref{fig:VI_ACC_Unstable}, the voltages and currents have an oscillation with a frequency around 8.5 Hz and are increasing exponentially. 

One can assume that the instability is resulting in (i) imperfect control tuning and modeling of the HVDC system and wind farms or (ii) interaction between the controls of wind farms and HVDC system. The first assumption can be checked by disconnecting the wind farms from the HVDC system and simulation can be carried out separately by connecting a simple $R-L$ load or constant power load (CPL) with the HVDC system and an ac grid with the wind farms. Thus, a CPL with the same rated power of the wind farm has been connected to the HVDC system and time domain simulation has been carried out. The system is found to be stable from the time domain simulation; therefore the HVDC system is stable itself for this tuning. Now the wind farm has been connected to a strong ac grid and the time domain simulation confirms that the wind farms operate stably without the HVDC system. Thus, the instability is resulting from the control interaction of the HVDC rectifier VSC-R and wind power inverter WECS-I. 

\begin{figure}[!t]
\centering
\includegraphics[width=.40\textwidth]{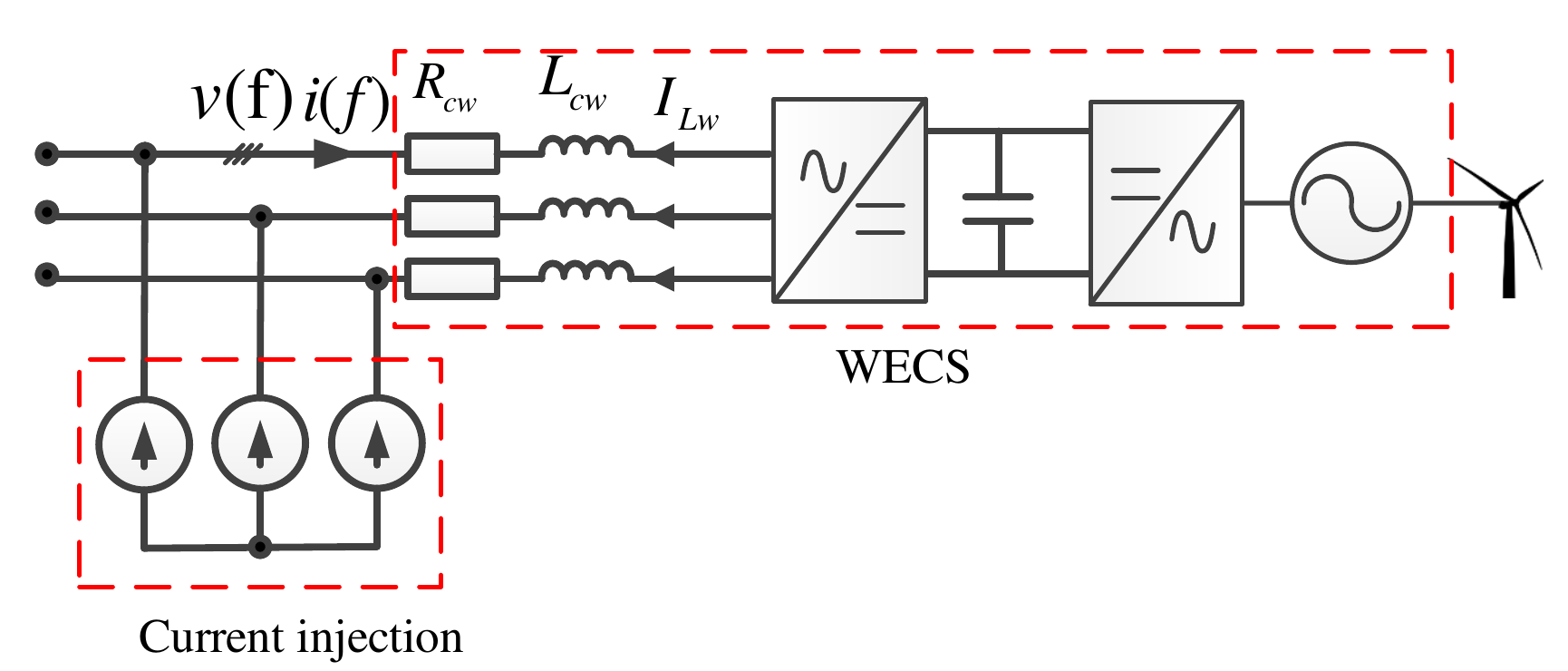}
\caption{Impedance measurement set up from the wind turbine generator.}
\label{fig:WECS_Sys_I_inj}
\end{figure}

\begin{figure}[t!]
\centering
\includegraphics[width=.50\textwidth]{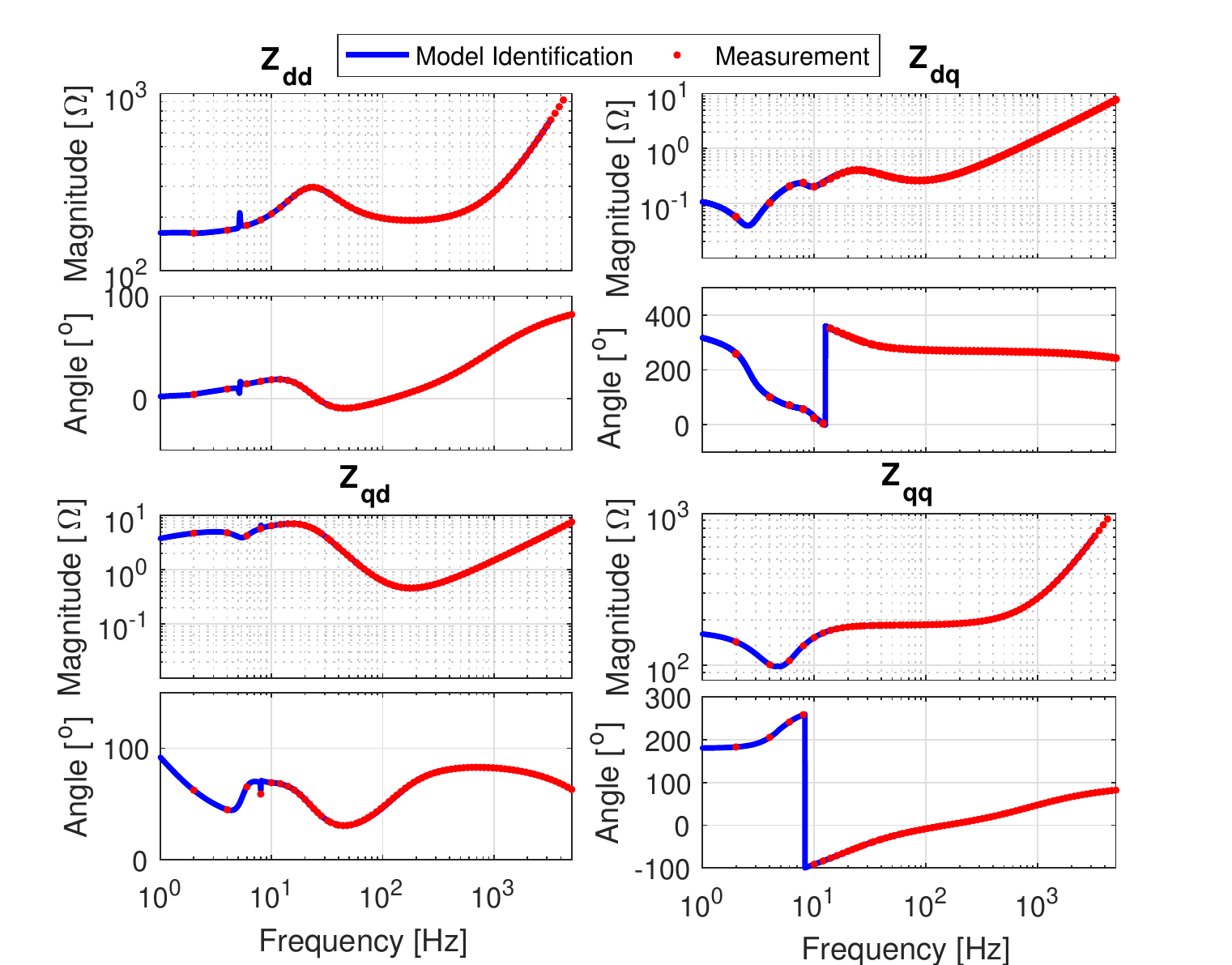}
\caption{Impedance frequency responses of the WECS (Solid line is model identification and the dots are from measurement).}
\label{fig:Imp_MID}
\end{figure}

\section{Proposed Impedance-based Stability Analysis}

\subsection{Identification of Impedance Model of WECS-I}
 The impedance-based method has been adopted to assess the stability of the interconnected system. In order to assess the stability based on the impedance method, a continuous transfer function of the impedance is required. An analytical transfer function of the impedance model of the WECS inverter can not be obtained since, the controller parameters are unknown. Thus, the following steps are taken to obtain the transfer function of the aggregated impedance of the wind farms. 
 
 The WTGs are assumed to be identical in structure and control from a vendor. One WTG has been connected to the main ac grid without the HVDC transmission line and the impedance frequency responses have been measured from 1 Hz to 5 kHz with 75 measurement points in dq-frame \cite{MAminJESTPE2016}. The measurement set-up is shown in Fig. \ref{fig:WECS_Sys_I_inj}. The measurement point has been selected randomly in logarithmic scale. More measurement point will give better approximation of the impedance model. A model identification technique has been used to estimate the transfer function for each element of the dq-domain impedance matrix as 
\begin{align}
Z_{_{WECS-I}}^{dq}=\frac{{{b}_{m}}{{s}^{m}}+{{b}_{m-1}}{{s}^{m-1}}+...+{{b}_{o}}}{{{a}_{n}}{{s}^{n}}+{{a}_{n-1}}{{s}^{n-1}}+...+{{a}_{o}}}
\end{align}

where $b_m$, $b_{m-1}$,...$b_o$, $a_n$, $a_{n-1}$,...$a_o$ are constant coefficients and $m$ and $n$ are the order of Zero and Pole of the impedance model.
The order of the transfer function is selected (5 in this case) such that the error between the measurement and the model identification is less than 0.10$\%$. Fig. \ref{fig:Imp_MID} shows the impedance frequency responses from the model identification with the measured impedance by point-by-point simulation. As can be seen, the obtained transfer function of the impedance model has very good agreement with measured impedance frequency responses with both magnitude and phase.

 \begin{figure}[!t]
\centering
\includegraphics[width=.30\textwidth]{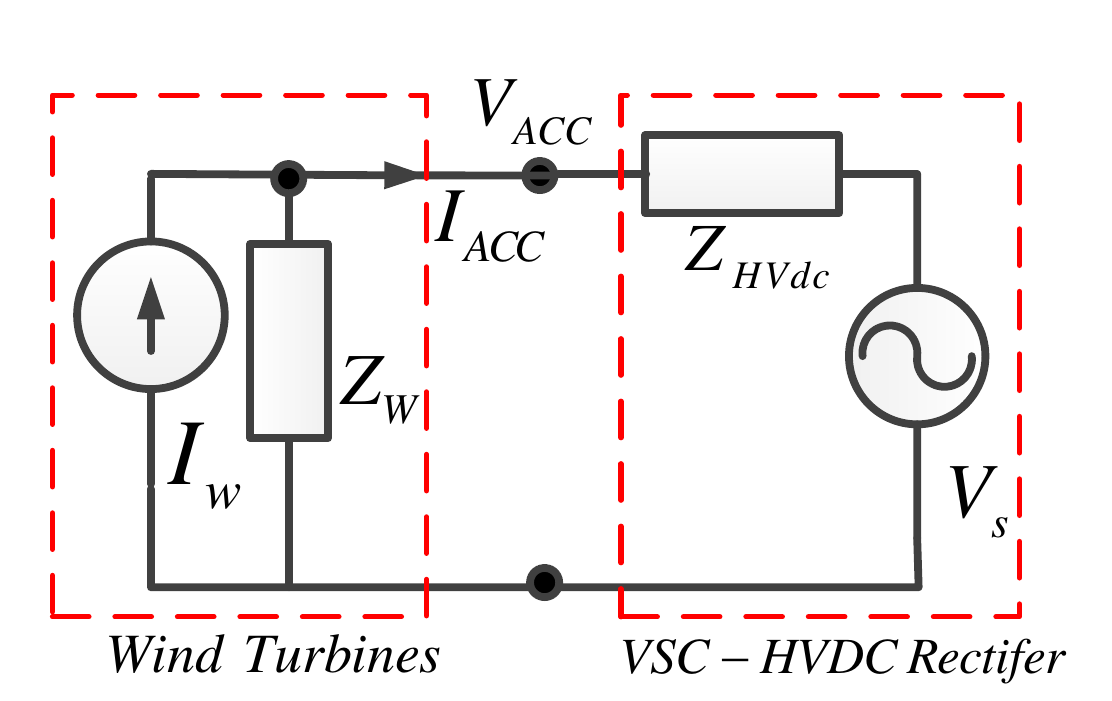}
\caption{Impedance based equivalent model of offshore ac grid system}
\label{fig:SmallSignalModel}
\end{figure}
 
\subsection{Stability Analysis}
The small-signal impedance model of interconnected system is shown in Fig. \ref{fig:SmallSignalModel} where $Z_{HVDC}$ and $Z_{W}$ are total impedance of the HVDC system and the wind farms from the offshore ACC point, respectively. The subscript for the $dq$ has been dropped in the equivalent circuit. $Z_{HVDC}$ is found together with the parallel connection of the HVDC rectifier capacitive filter impedance and series connection of transformer impedance and can be given by  
\begin{align}
\label{eqn:zHVDCdq}
{{Z}_{HVDC}^{dq}}={{Z}_{T,dq}}+{{\left( Z_{VSC-R}^{-1}+Z_{Cf,dq}^{-1} \right)}^{-1}}
\end{align}
where the analytical impedance model of the HVDC rectifier has been taken from \cite{MAminJESTPE2016} and is given by (\ref{zdqa}).  
 \begin{align}
\label{zdqa}
Z_{VSC-R}^{dq}=-\left({I-{{V}_{dc}}{{G}_{PWM}}\left( I-{{G}_{ccR}}{{G}_{vR}} \right)}\right)^{-1}. \nonumber \\
\left({{{Z}_{0R}}+{{V}_{dc}}{{G}_{PWM}}\left( {{G}_{ccR}}+{{G}_{delR}} \right)}\right)
\end{align}
The impedance of the wind farms from the ACC point can be given by
  \begin{align}
  \label{eqn:zwaccdq}
  {{Z}_{W}^{dq}}=\frac{1}{n}( {{Z}_{cable}}+{{Z}_{T,dq}}+{{( Z_{WECS-I}^{-1}+Z_{Cf,dq}^{-1} )}^{-1}} )
  \end{align}
where n is the number of wind farms. To obtain aggregated equivalent impedance of the wind farms, it has been assumed that all the WTGs are operating in the same steady-state point.

  Based on this representation in Fig. \ref{fig:SmallSignalModel}, the response of the ACC bus voltage can be written by (\ref{eqn:impr}).  
\begin{align}
\label{eqn:impr}
{{V}_{ACC}}(s)=\left( {{V}_{s}}(s)+{{Z}_{HVDC}^{dq}}(s)I(s) \right) 
\left( I+\frac{{{Z}_{HVDC}^{dq}}(s)}{{{Z}_{W}^{dq}}(s)} \right)^{-1}
\end{align}

For system stability studies, it is assumed that 
\begin{enumerate}
\item{The ac voltage of VSC-R is always stable when unloaded; and }
 \item{The wind farms current is stable when it is connected to a stable source.}
\end{enumerate}
Therefore, the stability of the interconnected system depends on the second term of right-hand side of (\ref{eqn:impr}) and the ACC bus voltage will be stable if and only if the impedance ratio matrix,  $(Z_{HVDC}^{dq}(s))(Z_{W}^{dq}(s))^{-1}$ which can be defined as the minor loop gain of feedback control system as  
\begin{align}
\label{eqn:mlg}
G(s)H(s)=\left({{{Z}_{HVDC}^{dq}}(s)}\right)\left({{{Z}_{W}^{dq}}(s)}\right)^{-1}
\end{align}
meets the Generalized Nyquist Stability Criterion (GNC) \cite{RDMiddlebrook1976}, \cite{JianSun2011}.

The stability of the system depends on the multi-input multi-output (MIMO) Nyquist variables since the impedance model is a 2x2 matrix and it is therefore necessary to include all the elements of matrix G(s)H(s) as described in \cite{RBurgosECCE2010}. 

\begin{figure}[!t]
\centering
\includegraphics[width=.40\textwidth]{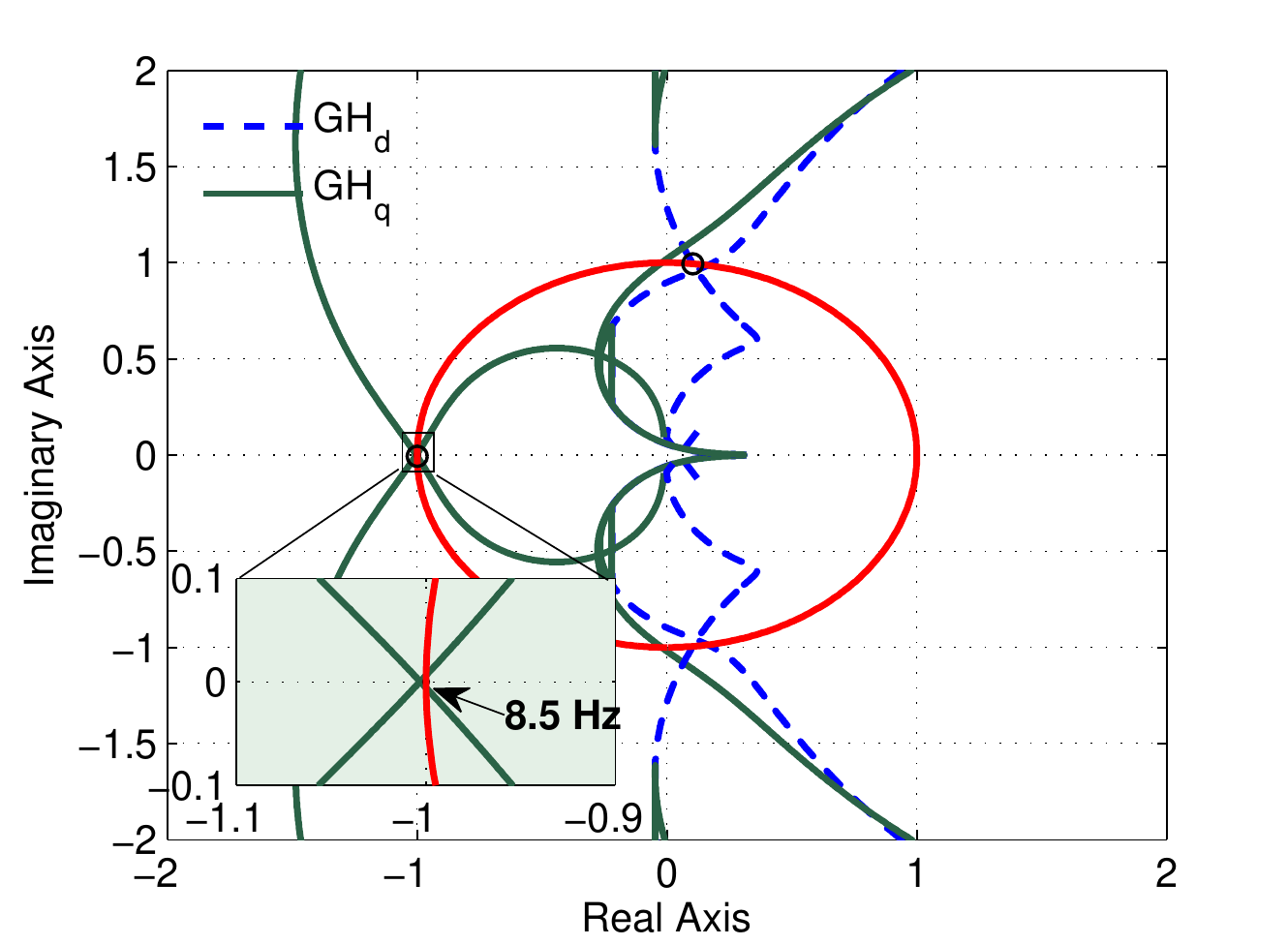}
\caption{Unstable: Nyquist plots of minor-loop gain.}
\label{fig:Nyquist_unstable}
\end{figure}

The $RLC$ parameter of the filter, transformer and sub-sea cable of the interconnected system are assumed to be known. Hence, the stability of the entire system can be effectively determined before connecting to the main ac grid since all the parameters are now available for stability analysis.
Fig. \ref{fig:Nyquist_unstable} shows the frequency domain stability analysis results for the simulation presented in previous section. As can be seen in Fig. \ref{fig:Nyquist_unstable}, the q-axis dominated Nyquist plot encircles the point (-1, j0) at frequency 8.5 Hz, the system becomes unstable. 

\begin{figure}[!t]
\centering
\includegraphics[width=.40\textwidth]{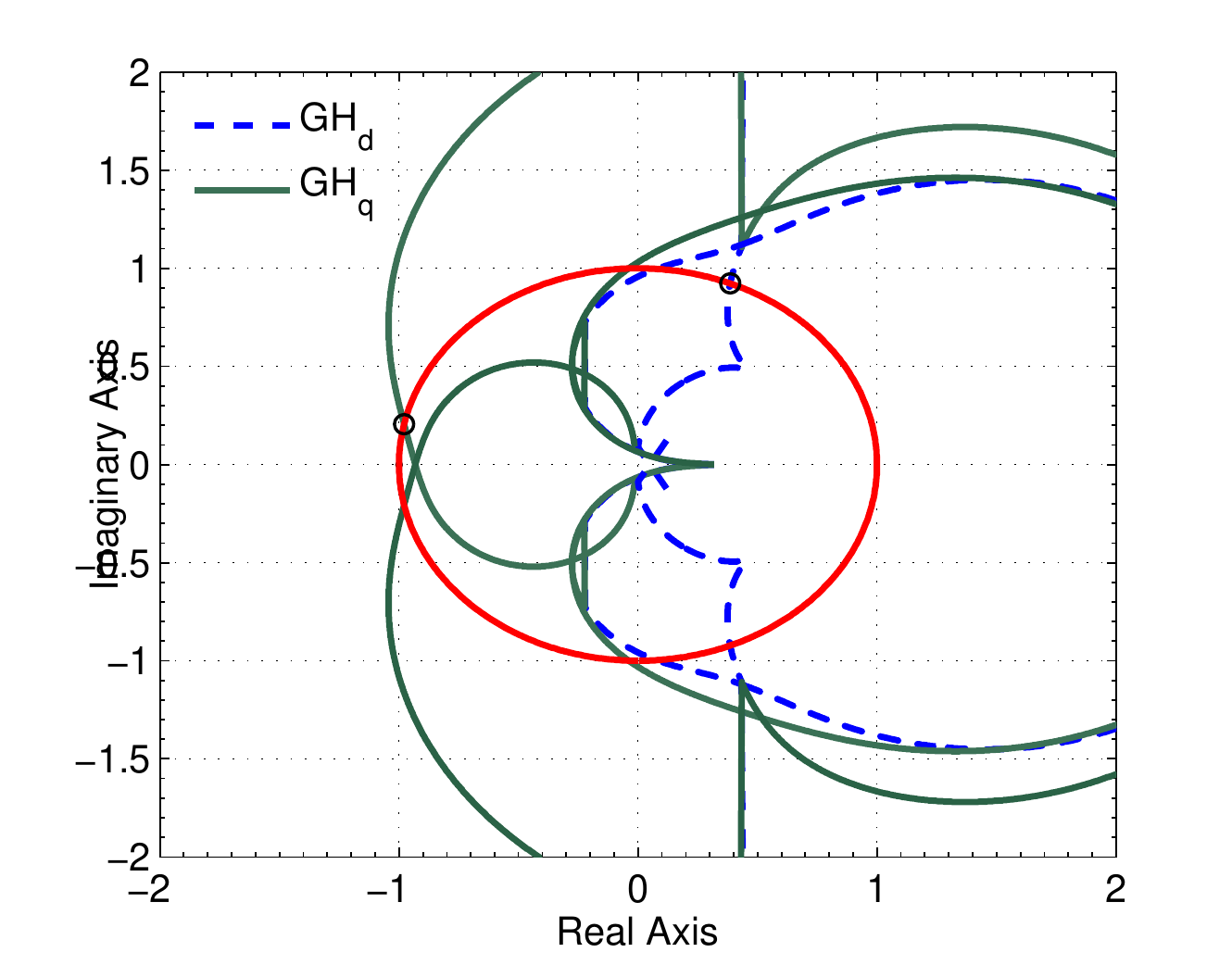}
\caption{Stable: Nyquist plots of minor-loop gain.}
\label{fig:Nyquist_stable}
\end{figure}

\begin{figure}[!t]
\centering
\includegraphics[width=.50\textwidth]{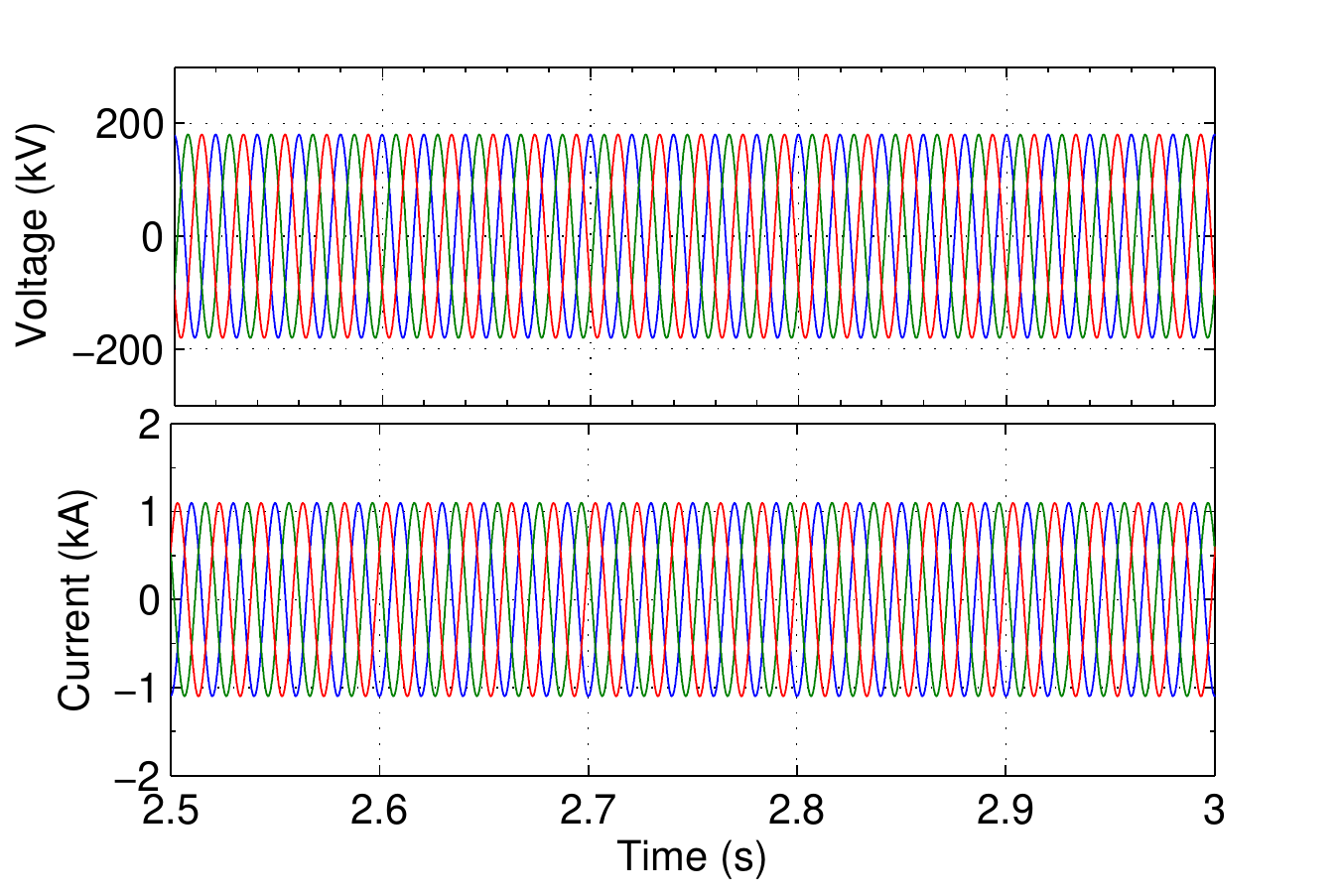}
\caption{Stable: Three phase voltages and currents at ACC bus.}
\label{fig:VI_ACC_Stable}
\end{figure}

The WTGs are assumed to be a black box and we do not know what to change in the WTGs to mitigate the instability causing by interaction of the controllers of the interconnected areas. According to \cite{HanchaoLiu2014IEEEJESTPE}, \cite{MAminJESTPE2016}, the performance of the system can be improved by increasing the ac voltage control-loop bandwidth of the HVDC rectifier. The ac voltage controller bandwidth of VSC-R was five times less than the inner-loop current controller bandwidth. A standard practice for the converter designer is that the outer-loop bandwidth should be three to ten times smaller than the inner control loop \cite{LHarneforsIEEETIE2007}. Therefore, the bandwidth of the ac voltage control loop has been increased. The ac voltage controller gain is chosen $H_{vac}(s)=0.2+40/s$ in pu with 58 degrees phase margin at 132 Hz crossover frequency while previously the crossover frequency was 80 Hz. Hence, the bandwidth of the ac voltage controller becomes three times smaller than the inner current control loop. Fig. \ref{fig:Nyquist_stable} shows the frequency domain stability analysis results for this tuning. As can be seen, the Nyquist plots do not encircle the point (-1, j0), hence the system will operate stably.
A time domain simulation has been carried out for the new control tuning of the ac voltage control loop while other control-loops bandwidth remain the same. Fig. \ref{fig:VI_ACC_Stable} shows the time domain responses from the simulation. The system operates stably for this tuning. 
Thus, the proposed stability analysis method based on the aggregated impedance predicts the stability of the interconnected system of the wind farms and the HVDC system even though no information of the WTGs is known.

The impedance-based stability is small-signal stability analysis and is valid only for small-range of operating point; however, the wind turbines operate at various wind speed and provides variable output power, therefore the stability analysis has been carried out at various operating point by measuring the impedance for various operating points and checking the GNC.
Though, the interconnected system operates stably for this new tuning, it is not clear yet which controller is participating in the observed oscillation. 

\section{Extraction of Controllers' Parameter and Interaction Analysis}

\subsection{Extraction of Inverter Controllers' Parameters}

The controller interactions are the likely sources of these oscillations, therefore the understanding of the electrical oscillation between interconnected area at its source remain of crucial importance. The impedance-based stability method can effectively determine the source of the instability of the interconnected system, the various configurations and controllers in real life systems are not always known in the details due to confidentiality and industry secrecy.
When engineers design the interconnected areas of HVDC system and wind farms it might use the wind turbines from various vendors with different controller dynamics and the system designer needs the information about the control of the wind turbines inverter to make the interconnected system operate stably. This Section presents a method to reveal the internal dynamics of WECS-I from the measurement data while the WTGs are considered to be a black box.  

The measured diagonal elements of the WECS-I impedance can be represented by a transfer function using system identification technique as 
 \begin{subequations}
\label{eqn:zdd_est}
\begin{align}
{{Z}_{dd\_est}}=\frac{{{b}_{d5}}{{s}^{5}}+{{b}_{d4}}{{s}^{4}}+...+{{b}_{do}}}{{{a}_{d4}}{{s}^{4}}+{{a}_{d3}}{{s}^{3}}+...+{{a}_{do}}}
\end{align}
\begin{align}
{{Z}_{qq\_est}}=\frac{{{b}_{q5}}{{s}^{5}}+{{b}_{q4}}{{s}^{4}}+...+{{b}_{qo}}}{{{a}_{q4}}{{s}^{4}}+{{a}_{q3}}{{s}^{3}}+...+{{a}_{qo}}}.
\end{align}
\end{subequations}

From \cite{MAminJESTPE2016}, the analytical impedance model of the WECS inverter for the dc voltage and reactive power control can be given by   
\begin{align}
\label{eqn:zdqw}
Z_{WECS-I}^{dq}=\left({I-{{V}_{dc}}G_{A}^{-1}{{G}_{C}}+{{G}_{D}}{{G}_{vd}}G_{A}^{-1}{{G}_{C}}}\right)^{-1}. \nonumber \\
\left({-{{Z}_{0w}}-{{G}_{D}}{{G}_{vi}}+({{V}_{dc}I}-{{G}_{D}}{{G}_{vd}})(G_{A}^{-1}{{G}_{B}})}\right).
\end{align}

The impedance model in (\ref{eqn:zdqw}) is in dq-frame and a 2x2 matrix. 
The diagonal elements of (\ref{eqn:zdqw}) can be written as a function of control-loop gain as
\begin{subequations}
\label{eqn:zddw}
\begin{align}
Z_{WECS-I}^{dd}=\frac{Z_0+\frac{{{D}_{d}}}{s{{C}_{dc}}}{{D}_{d}}+\left( {{V}_{DC}}-\frac{{{D}_{d}}}{s{{C}_{dc}}}{{I}_{d}} \right)\frac{{{Z}_{o}}{{\psi }_{n}}}{{{\psi }_{d}}}}{1-\left( {{V}_{DC}}-\frac{{{D}_{d}}}{s{{C}_{dc}}}{{I}_{d}} \right)\frac{{{H}_{pwm}}}{{{\psi }_{d}}}}
\end{align}
\begin{align}
Z_{WECS-I}^{qq}= \frac{{{Z}_{0}}+\frac{D_{q}^{2}}{s{{C}_{dc}}}+{{G}_{cc-ol}}{{Z}_{0}}}{1-{{V}_{DC}}{{H}_{pwm}}-{{G}_{PLL}}{\psi }_{PLL}}
\end{align}
\end{subequations}
where
\begin{align}
  & {{\psi }_{n}}=\left( {{D}_{d}}{{G}_{vdc-ol}}(1+{{G}_{cc-ol}})+{{G}_{cc-ol}} \right)  \nonumber \\
& {{\psi }_{d}}=1+{{I}_{d}}{{Z}_{o}}{{G}_{vdc-ol}}(1+{{G}_{cc-ol}})  \nonumber \\
&{{\psi }_{PLL}}={{G}_{cc-ol}}{{Z}_{0}}{{I}_{d}}-{{V}_{DC}}{{H}_{pwm}}{{V}_{d}}+{{V}_{DC}}{{D}_{d}}. \nonumber 
\end{align}
and $G_{cc-ol}$ and $G_{vdc-ol}$ are the open-loop transfer function of the current and dc voltage control-loop, respectively, $G_{PLL}$ is the close-loop gain of the PLL and $Z_0=R_{cw}+sL_{cw}$.

\begin{figure}[t!]
\centering
\includegraphics[width=.50\textwidth]{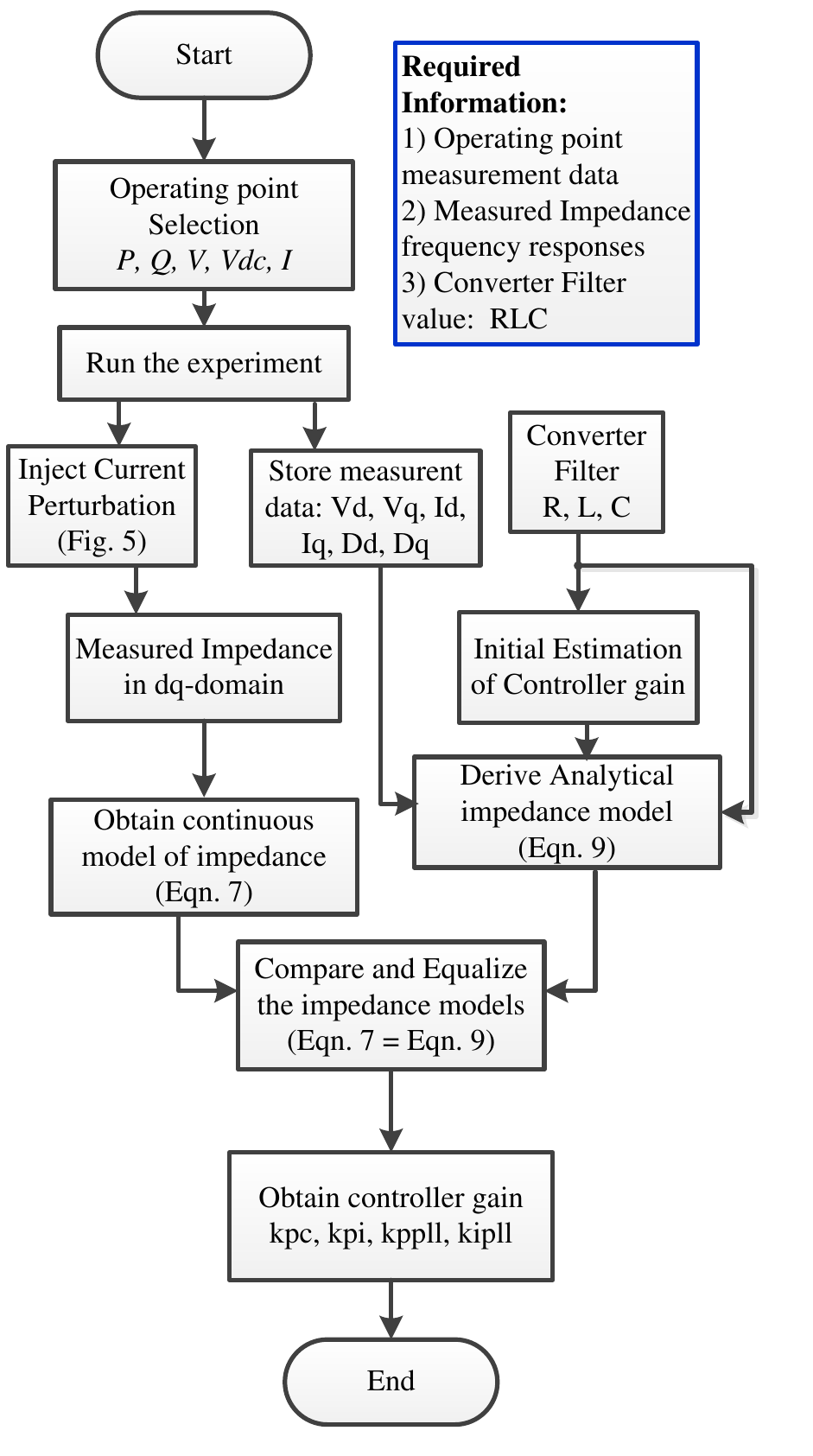}
\caption{ Illustration of controller gain extraction method from measured impedance.}
\label{fig:FlowChart_BWR}
\end{figure}

The analytical impedance model of (\ref{eqn:zddw}) can replaced by measured impedance model as 
 \begin{subequations}
\label{eqn:zddw_est}
\begin{align}
{{Z}_{dd-est}}=\frac{{{Z}_{0}}+\frac{{{D}_{d}}}{s{{C}_{dc}}}{{D}_{d}}+\left( {{V}_{DC}}-\frac{{{D}_{d}}}{s{{C}_{dc}}}{{I}_{d}} \right)\frac{{{Z}_{o}}{{\psi }_{n}}}{{{\psi }_{d}}}}{1-\left( {{V}_{DC}}-\frac{{{D}_{d}}}{s{{C}_{dc}}}{{I}_{d}} \right)\frac{{{H}_{pwm}}}{{{\psi }_{d}}}}
\end{align}
\begin{align}
{{Z}_{qq-est}}= \frac{{{Z}_{0}}+\frac{D_{q}^{2}}{s{{C}_{dc}}}+{{G}_{cc-ol}}{{Z}_{0}}}{1-{{V}_{DC}}{{H}_{pwm}}-{{G}_{PLL}}{\psi }_{PLL}}.
\end{align}
\end{subequations}
In (\ref{eqn:zddw_est}), all the elements are known from measurement data except the proportional and integral gain of the current controller, dc voltage controller and the PLL. We have two equations with six unknown variables. In order to solve these equations,  (\ref{eqn:zddw_est}) is represented in $j\omega$-domain and the required number of equations can be obtained at different frequencies from measurement data and by separating them in real and imaginary part. Now solving those equations by iteration, the controller gains are obtained.
The bandwidth extraction method is presented through a flow chart as shown in Fig. \ref{fig:FlowChart_BWR}. The method is presented for the decoupled d-q frame control, however this procedure can be applied to other type of control of VSC to extract the controller dynamics.  In order to identify controller dynamics for other type of converter control, first it is necessary to derive the analytical impedance model and then, the method can be applied to extract the controller gain/bandwidth.

\begin{figure}[t!]
\centering
\includegraphics[width=.50\textwidth]{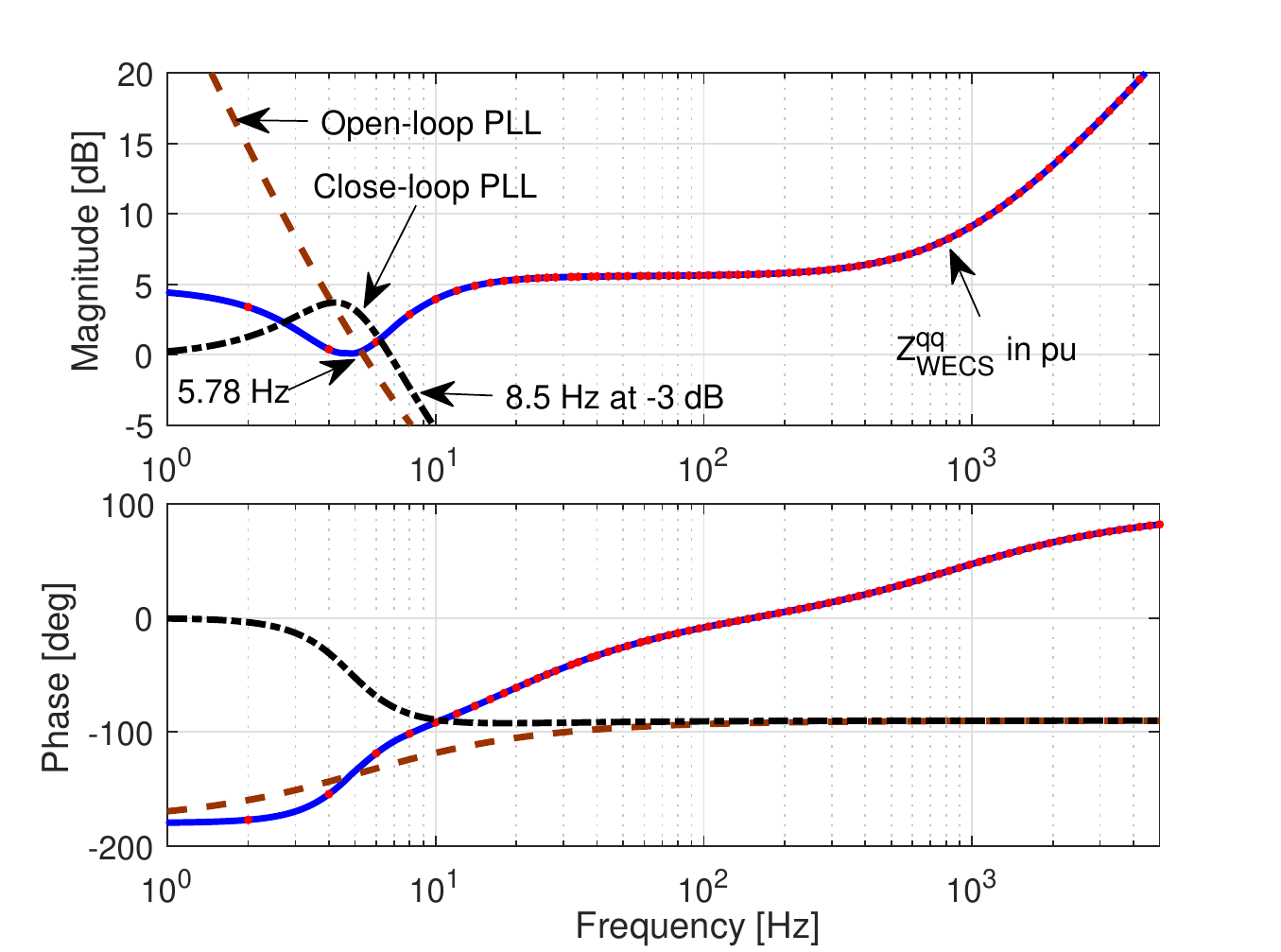}
\caption{ Control-loop gain of the PLL and the q-axis impedance of the WECS-I (Solid line is model identification and the dots are from simulation).}
\label{fig:hpllImp}
\end{figure}
\begin{figure}[!t]
\centering
\includegraphics[width=.50\textwidth]{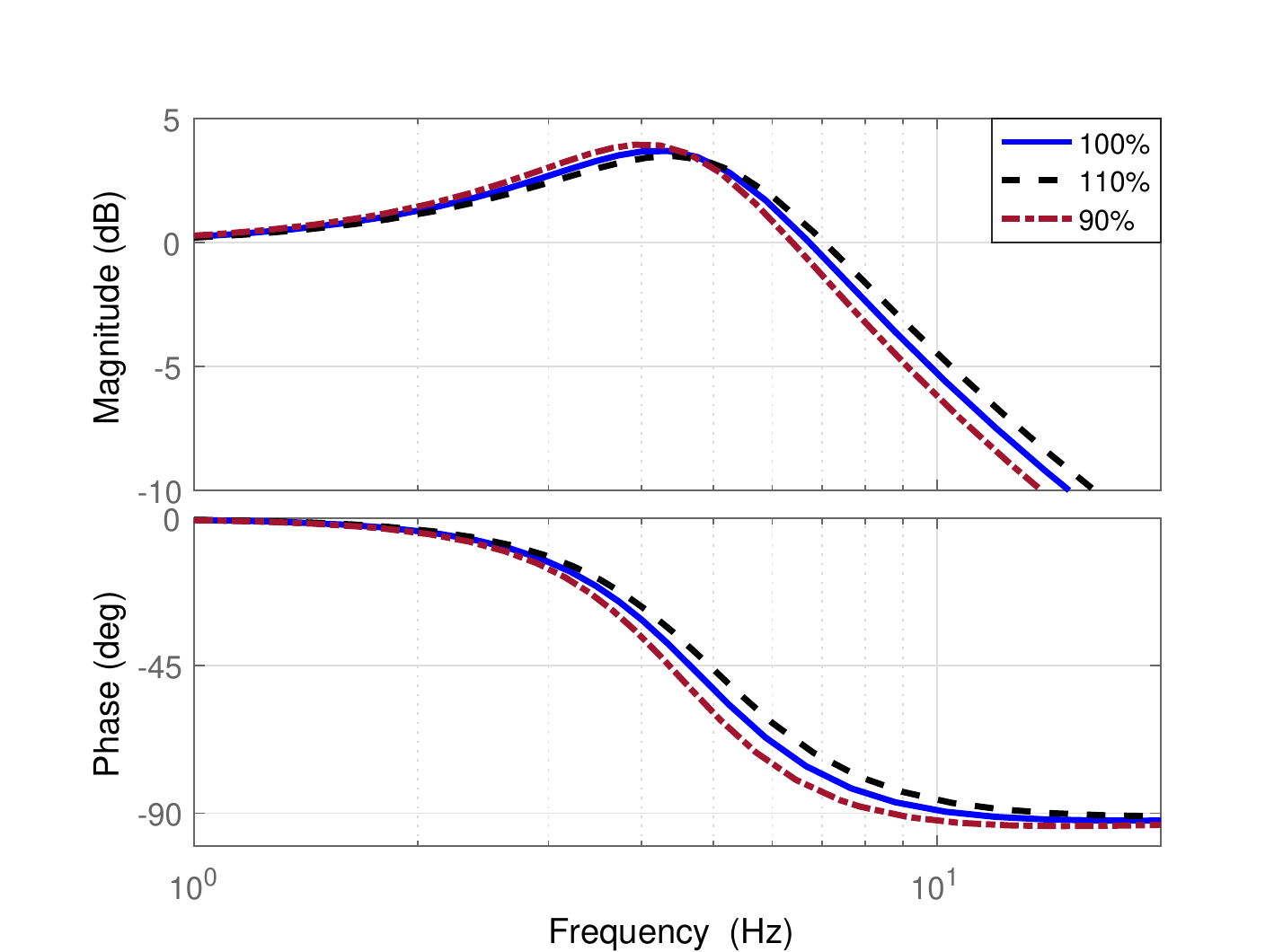}
\caption{Close-loop frequency response of the PLL from estimation.}
\label{fig:G_PLL}
\end{figure}

The bandwidth information of the PLL is extracted from the qq-axis impedance, since PLL has the most dominant impact on the qq-axis impedance. Fig. \ref{fig:hpllImp} shows the control-loop gain of the PLL where PLL open-loop gain has a phase margin 47 degrees at 5.78 Hz and the close-loop has a bandwidth around 8.5 Hz. Moreover, Fig. \ref{fig:hpllImp} shows the q-axis impedance of the wind farm inverter from the model identification and numerical simulation. As can be seen, the impedance has a resonance at low frequency around 5 Hz which indicates crossover frequency of the PLL open-loop gain. The bandwidth is calculated at -3dB magnitude, therefore if we move forward 3 dB more, the frequency is found 8.5 Hz which is the close-loop bandwidth of the PLL. 
This extraction method will give the system designer a view of the control bandwidth range which will be useful for the design of the entire system to avoid the control interaction.

\begin{figure}[!t]
\centering
\includegraphics[width=.50\textwidth]{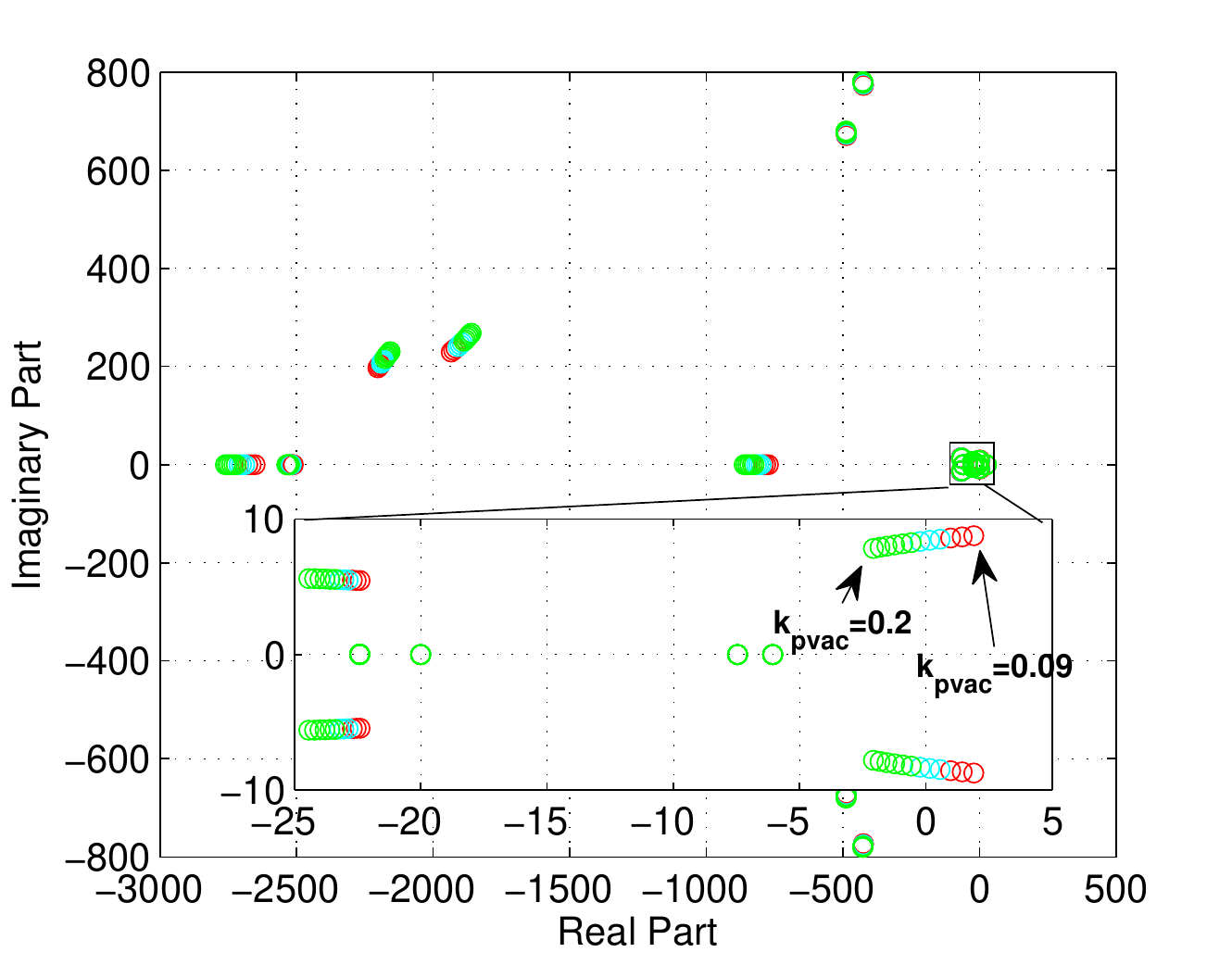}
\caption{Trajectory of eigenvalue for a change of proportional gain of ac voltage controller.}
\label{fig:Eigenvalue_kpvac}
\end{figure}

Fig. \ref{fig:G_PLL} shows the extracted proportional and integral gain of the PLL from the measurement data for three different cases. Since, the extracted controller gain might not be exact, two other cases with 10$\%$ error are presented for a comparison. For $10\%$ higher gain estimation case, the PLL loop bandwidth is found 8.9 Hz (see 110\% in Fig. \ref{fig:G_PLL}) and for $10\%$ lower estimation case, the PLL loop bandwidth is found 7.9 Hz (see 90\% in Fig. \ref{fig:G_PLL}) while the original bandwidth is 8.5 Hz. The estimated controller dynamics from measurement has good accuracy. The method effectively reveals the internal dynamics of the converters.

\begin{figure*}[!t]
\centering
\includegraphics[width=.80\textwidth]{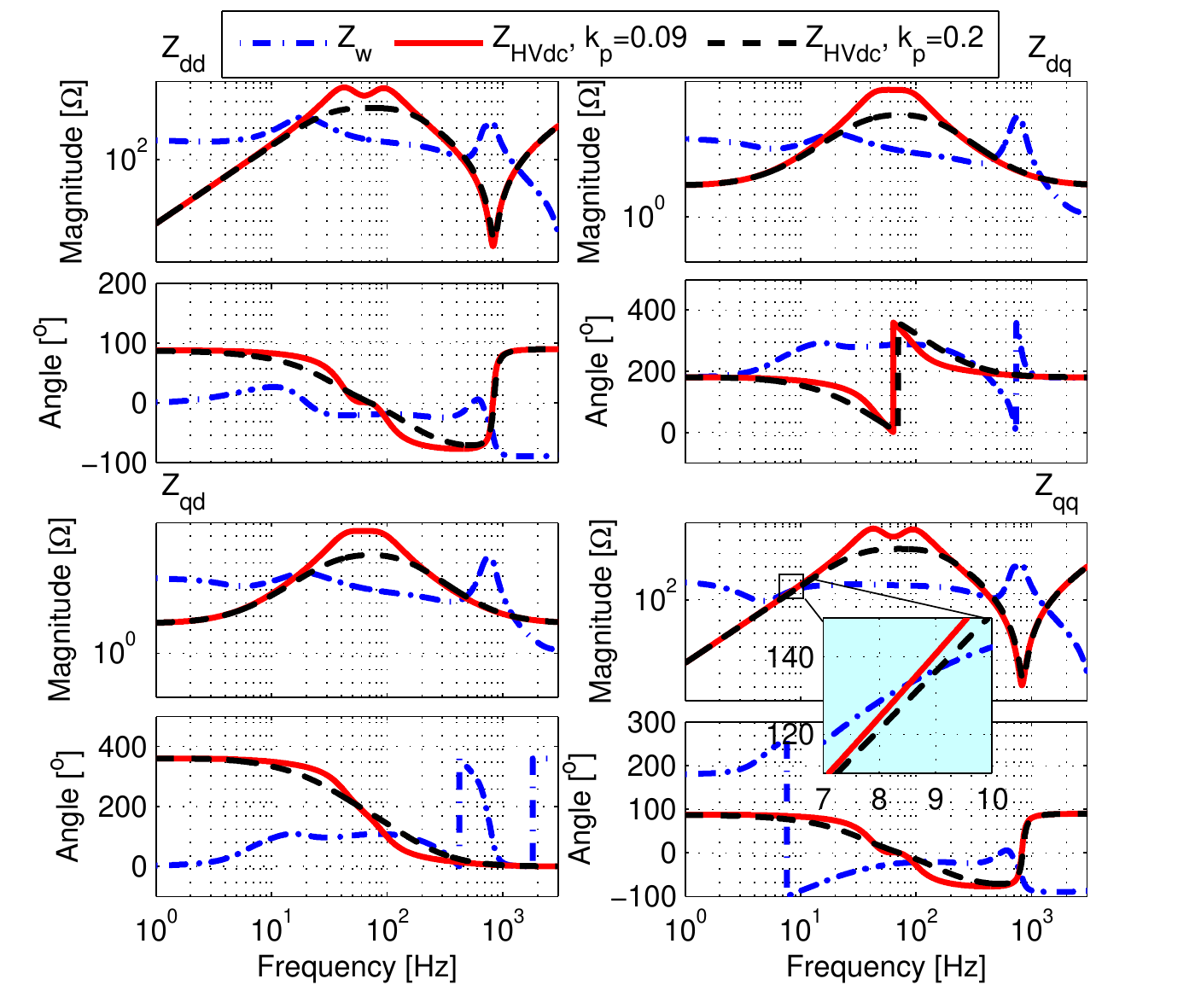}
\caption{Impedance frequency responses of the wind farms and the HVDC system from ac collection point for two cases of ac voltage controller tuning (Dash-dot line is for the wind farms, solid line and dash line is for the HVDC system for two control tuning of the ac voltage control).}
\label{fig:Imp_vacVar}
\end{figure*} 

\subsection{Interaction Analysis}
In order identify which controller is participating in this oscillation, we focus on the diagonal elements of the VSC-R and WECS-I impedance model since the diagonal elements are dominating along the entire Nyquist path. 

The diagonal elements of the VSC-R impedance from (\ref{zdqa}) can be written by

\begin{align}
\label{eqn:zddqqvscr}
{{Z}_{dig,VSC-R}}=\frac{{V_{dc0}{H}_{pwm}}{{H}_{i}}+{{R}_{c}+sL_c}}{1-{{H}_{pwm}}+{V_{dc0}{H}_{pwm}}{{H}_{i}}{{H}_{vac}}}
\end{align}
where ${{Z}_{dig,VSC-R}}={{Z}_{VSC-R}^{dd}}={{Z}_{VSC-R}^{qq}}$. Neglecting PWM delay, (\ref{eqn:zddqqvscr}) can be written as a function of current control-loop as
\begin{align}
{{Z}_{dig,VSC-R}}=\frac{1}{{{G}_{cc-cl,VSC-R}}{{H}_{vac}}}
\end{align}
where ${G}_{cc-cl,VSC-R}$ is the close-loop transfer function of current controller and can be written as
 \begin{align}
{{G}_{cc-cl,VSC-R}}=\frac{V_{dc0}{{H}_{pwm}{H}_{i,VSC-R}}/\left( {{R}_{c}}+s{{L}_{c}} \right)}{1+{V_{dc0}{H}_{pwm}{H}_{i,VSC-R}}/\left( {{R}_{c}}+s{{L}_{c}} \right)}.
\end{align}

Including the filter capacitor, the diagonal elements can be given by
\begin{align}
{{Z}_{dia,HVDC}}=\frac{{{Z}_{dig,VSC-R}}}{1+s{{C}_{f}}{{Z}_{dig,VSC-R}}}
\end{align}
which can be expressed as a function of ac voltage control-loop as
\begin{align}
\label{eqn:zdiagHVDC}
{{Z}_{dig,HVDC}}=\frac{1}{s{{C}_{f}}({{G}_{vac-ol}}+1)}
\end{align}
where ${G}_{vac-ol}$ is the open-loop transfer function of the ac voltage control-loop and can be given by
\begin{align}
{{G}_{vac-ol}}={{H}_{vac}}{{G}_{cc-cl}}\frac{1}{s{{C}_{f}}}.
\end{align}
From (\ref{eqn:zdiagHVDC}), we see that HVDC rectifier impedance depends on the open-loop transfer function of the ac voltage controller. The higher control bandwidth of the ac voltage control-loop is meaning the lower impedance magnitude of the HVDC converter. Moreover, the d- and q-axis impedances are equal in magnitude and do not depend on the operating point.

The diagonal elements of WECS-I impedance in (\ref{eqn:zddw}) indicate that the d-axis impedance of the WECS-I depends on the outer-loop dc voltage controller, inner-loop current controller and operating point, and the q-axis impedance depends on the close-loop PLL bandwidth, inner-loop current controller and operating point. 

\begin{figure*}[!t]
\centering
\includegraphics[width=.80\textwidth]{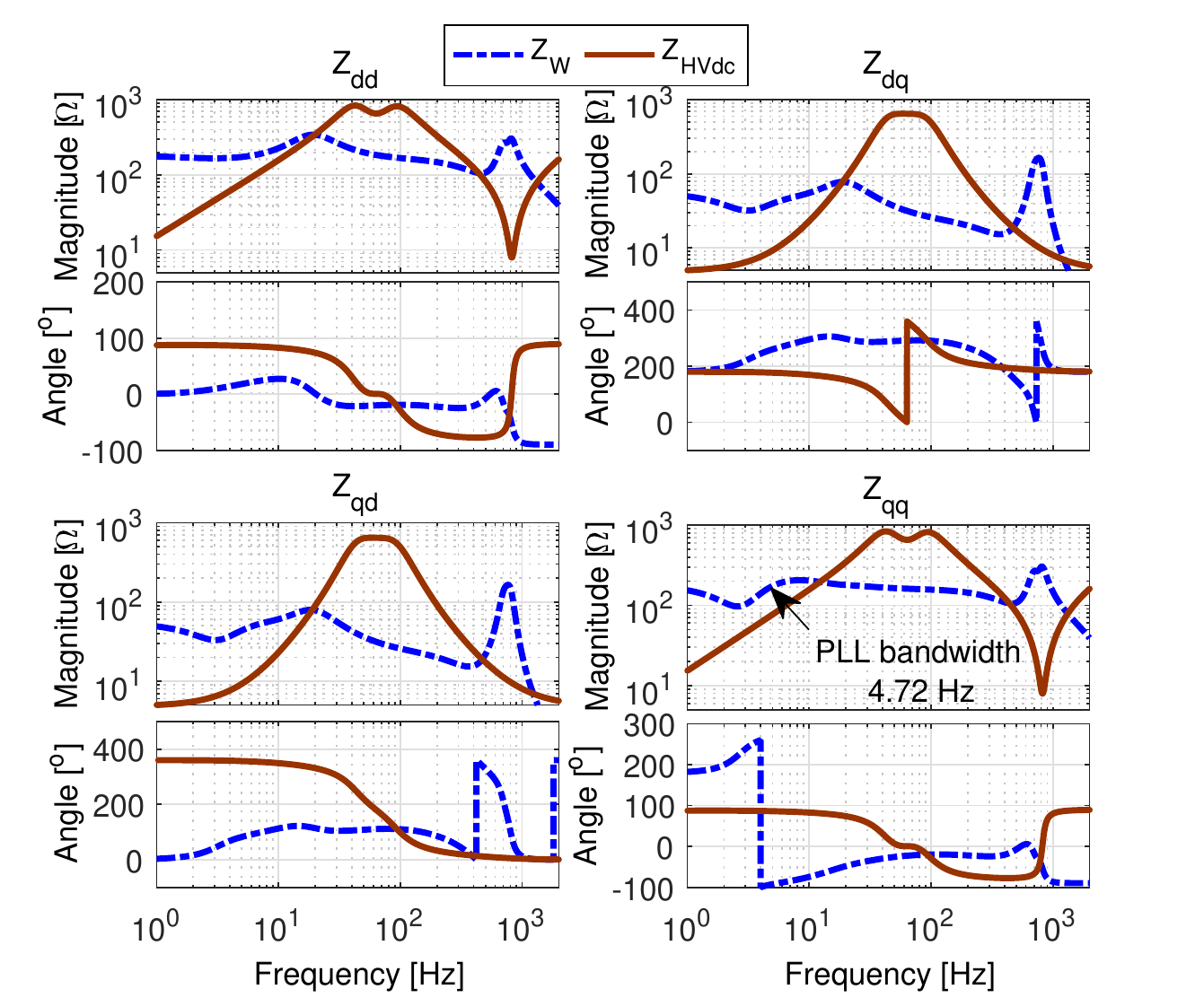}
\caption{Impedance frequency responses of the wind farms and the HVDC system from ac collection point for retuning the PLL (Solid line is the HVDC and dash line is for the wind farms).}
\label{fig:Imp_pll}
\end{figure*}

The passive components such as inductor, capacitor and resistor has an equal impact on both the d- and q-axis impedance, however the controllers in the wind farm inverters are different since d-axis has outer-loop dc voltage control while the q-axis has PLL synchronization loop. Now we can observe the Nyquist plots of the minor-loop gain shown in Fig. \ref{fig:Nyquist_unstable}. As can be seen, the d-axis impedance dominated Nyquist plot has sufficient phase margin and is far from the stability marginal point while the Nyquist plot of the q-axis dominated impedance encircles the point (-1, j0). Thus the controllers in the q-axis are interacting resulting in instability in the interconnected areas. Source HVDC rectifier's most outer control in the q-axis is ac voltage control-loop and wind farms inverter's most outer controller in the q-axis is the PLL. The controllers in VSC-R and WECS-I with the slowest bandwidth are dominating part in the impedance magnitude at low frequencies, therefore the ac voltage controller of the HVDC rectifier and the PLL of WECS-I are interacting resulting in instability.
  
To confirm the instability resulting the interaction of the ac voltage controller of the HVDC rectifier and WECS inverter, a state-space model has been derived for the system and the participation factor analysis has been carried out for the unstable eigenvalues to identify the contribution of the states.
For simplification of analytical state-space modeling, the turbine mechanical system including the generator and generator side VSC is replaced by a power source \cite{HanchaoLiu2014IEEEJESTPE}. Fig. \ref{fig:Eigenvalue_kpvac} shows the trajectory of the eigenvalues for a change of proportional gain of ac voltage controller. The system has an unstable complex conjugate eigenvalue at $1.9\pm j 2 \pi 8.75$ with a oscillation frequency of 8.75 Hz for $k_{pvac}=0.09$. This is the unstable case that we see in the time domain simulation shown in Fig. \ref{fig:VI_ACC_Unstable} and frequency domain analysis shown in Fig. \ref{fig:Nyquist_unstable}. From the participation factor analysis, it is found that the PLL and ac voltage controller are the most contributing states for the unstable eigenvalue.
The participation factor analysis proves that the instability is resulting in interaction of the ac voltage controller and the PLL which gives the same conclusion as the presented bandwidth extraction method from impedance-based method. 

\subsection{Mitigation Method}
Fig. \ref{fig:Imp_vacVar} shows the impedance frequency responses of the wind farm and the HVDC system from ac collection point. The HVDC rectifier impedances are shown for the two cases with proportional gain 0.09 and 0.2 of the ac voltage controller. Since the system becomes unstable only for the q-axis impedance, the q-axis impedance is of most interest. The q-axis impedance of the HVDC rectifier behaves as inductive up to a frequency around 40 Hz and it becomes a bandpass filter at frequency of the ac voltage control-loop bandwidth. However, the q-axis impedance of the wind farm behaves capacitive as up to the crossover frequency of the PLL-loop and shows  the characteristics of band reject filter. As can be seen in Fig. \ref{fig:Imp_vacVar}, the $Z_{qq}$ impedance of the wind farm intersects the HVDC impedance at 8.5 Hz and 9 Hz for ac voltage-controller proportional gain 0.09 and 0.2, respectively. From the close-loop transfer function of the PLL, the bandwidth is found 8.5 Hz. 
The system becomes unstable if the q-axis impedance magnitude of the HVDC system becomes larger than the wind farms impedance magnitude at frequency below the bandwidth of the PLL. 
In order to avoid the interaction between the ac voltage controller of the VSC-R and PLL of WECS-I, the controls of the both VSCs need to be re-tuned in such a way that the q-axis impedance magnitude of the HVDC system is kept lower than the impedance magnitude of wind farm at frequency of the PLL bandwidth.  

\begin{figure}[!t]
\centering
\includegraphics[width=.40\textwidth]{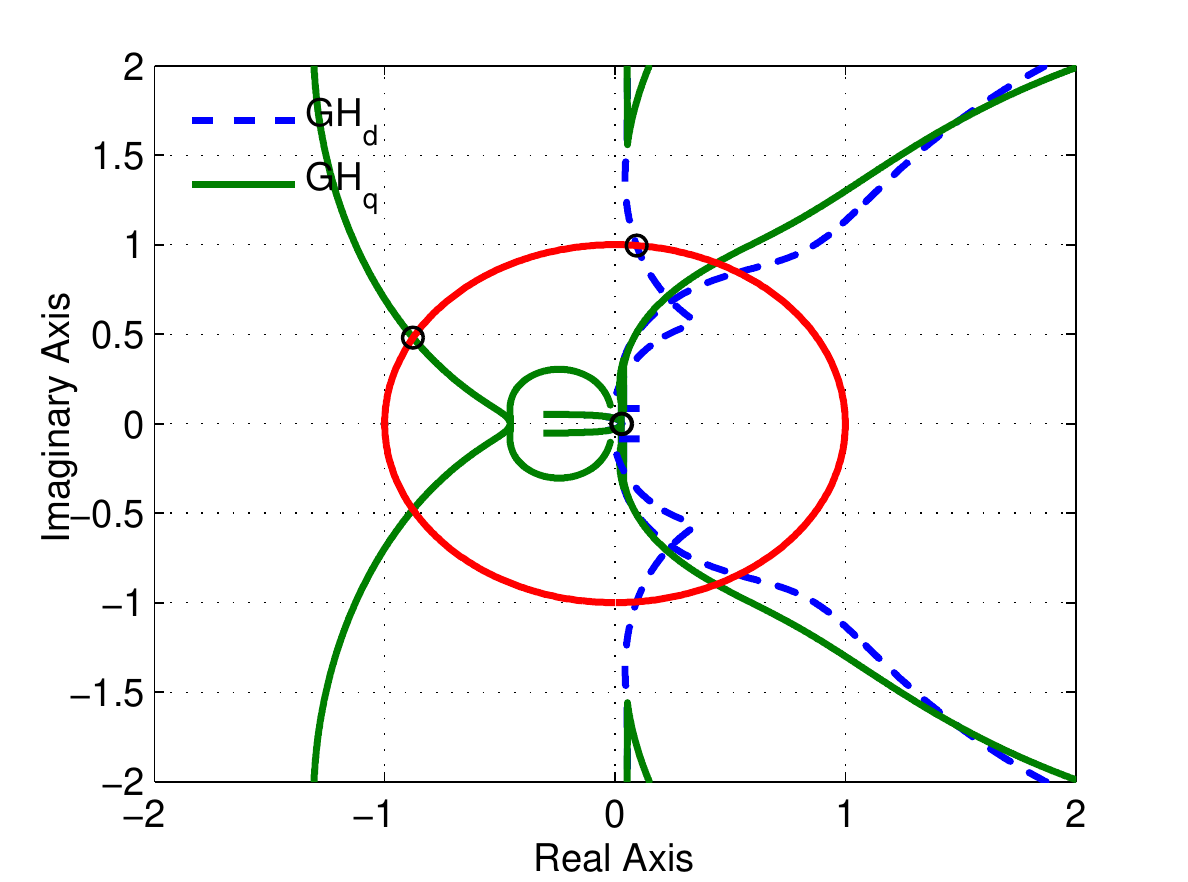}
\caption{Stable case: Nyquist plots of minor-loop gain for retuning of the PLL.}
\label{fig:Nyquist_stable_pll}
\end{figure}

A general rule is observed that if the PLL open-loop phase margin is more than 40 degrees and the close-loop bandwidth is 10 times smaller than the open-loop crossover frequency of the ac voltage controller, the system orerates stably.
Another example case is shown to prove this observation by retuning the PLL loop-gain at 3.17 Hz crossover frequency with 42 degrees phase margin and the PLL close-loop control bandwidth is 4.72 Hz for the unstable case of Fig. \ref{fig:VI_ACC_Unstable} and Fig. \ref{fig:Nyquist_unstable}. Fig. \ref{fig:Imp_pll} shows the impedance frequency responses of wind farms from ac collection point together with the HVDC impedance for new tuning of the PLL. As can be seen, the $Z_{qq}$ impedance magnitude of the wind farms does not intersect the HVDC impedance at frequency below the cutoff frequency of the PLL-loop. Fig. \ref{fig:Nyquist_stable_pll} shows the Nyquist plot of minor-loop gain. As can be seen, the Nyquist plots do not encircle the point (-1, j0), the system operates stably which has been confirmed by time domain simulation. 

The mitigation method presented for retuning the PLL is applicable for any grid tied inverter for example, the wind and solar power application. The q-axis impedance of the grid must not intersect the grid tied inverter impedance below the bandwidth of the PLL.

\section{Conclusion}
 
This paper proposes a method that enables to identify critical controllers' parameters from the the measurements of  frequency domain equivalent impedance (e.g. critical controller's bandwidth)  in an interconnected system.   
For doing that, the paper analyses the stability of an interconnected system of wind farms and HVDC transmission system. The impedance frequency responses of the wind farms and the HVDC system are measured at the ac collection point and it is shown how with this method it is possible to identify which controller in the interconnected system has a major impact in the observed system oscillations. 
A mitigation technique is proposed based on re-tuning of the corresponding critical controller bandwidth of the interconnected converters. A general observed rule is that the HVDC rectifier controller's bandwidth should be ten times higher than the control bandwidth of the wind power inverter's PLL.
The method suggested can reveal the internal controllers' dynamics of the wind farm from the measured impedance frequency responses combined with a general analytical expression of the impedance and an identified identical transfer function when no information about the controllers is provided by the vendors due to confidentiality and industry secrecy.
The method presented has potential immediate applicability in the wind industry based on the simplicity it offers to black/grey-box types of systems to guarantee stability of the interconnection, e. g. impedance measurement based, minimal required information of WECs detail design parameters, good accuracy.


%





\ifCLASSOPTIONcaptionsoff
  \newpage
\fi


\end{document}